## Title

General-Purpose Nonlinear Function Approximation via Linear Integrated Photonics

## Authors and Affiliation

Ayana Mizuno[1*], Isamu Takai[1], Makoto Nakai[1], Atsutaka Miyamichi[1], Ryuichi Konishi[2], Satoshi Sunada[3]

1. TOYOTA CENTRAL R&D LABS., INC., 41-1, Yokomichi, Nagakute, Aichi, 480-1192, Japan
2. Graduate School of Natural Science and Technology, Kanazawa University, Kakuma-machi, Kanazawa, Ishikawa, 920-1192, Japan
3. Faculty of Mechanical Engineering, Institute of Science and Engineering, Kanazawa University, Kakuma-machi, Kanazawa, Ishikawa, 920-1192, Japan

* Corresponding author: Ayana.Mizuno.zy@mosk.tytlabs.co.jp

## Abstract

Photonic computing has emerged as a promising platform for accelerating artificial intelligence workloads by enabling low-latency and energy-efficient linear operations such as vector–matrix multiplication. However, scalable on-chip high-order nonlinear processing remains challenging, limiting the functional versatility of current photonic hardware. Here, we present an optoelectronic approach for approximating high-order and high-dimensional nonlinear functions. The key to this approach lies in optical random Fourier feature mapping, which transforms nonlinear function evaluation into an equivalent linear computation. This approach enables nonlinear computing within a linear photonic framework, eliminating the need for complex optical nonlinear or active materials while preserving scalability and computational throughput in a simple silicon photonic circuit. We experimentally demonstrate a broad class of nonlinear functions, including tenth-order Legendre polynomials, computationally demanding special functions (Voigt, Fermi–Dirac, and Fresnel), neural-network activation functions, two-dimensional nonlinear functions, and a 10-dimensional softmax layer. This work establishes a general and scalable strategy for nonlinear computing in photonic integrated hardware and opens a pathway toward fully functional optical accelerators for next-generation computing systems.

## Introduction

Photonic computing has attracted growing interest as a hardware platform for accelerating computation beyond the limits of conventional electronics [1-5]. By exploiting the high bandwidth, low latency, and inherent parallelism of light, integrated photonics can implement linear operations efficiently [6-8], particularly vector–matrix multiplication [9-14]. These capabilities have driven significant progress in optical accelerators for artificial intelligence [9-15], signal processing [16-19], and scientific simulations [20-22].

Recent advances in silicon photonic platforms have demonstrated scalable and energy-efficient linear transformations, which highlight the maturity of photonic hardware for linear algebra operations. However, a wide range of computational tasks fundamentally rely on nonlinear operations. Nonlinear functions are indispensable in machine learning, nonlinear differential equation solvers, optimization, and signal processing [23-25]. Despite significant progress in linear photonic processors, the realization of an integrated platform capable of approximating a broad class of nonlinear functions remains a major challenge. This underscores the need for robust and flexible nonlinear photonic computing technologies to unlock broader application domains [26].

One major direction focuses on physical optical nonlinearities, including Kerr effects, carrier-induced refractive index changes, and semiconductor gain or absorption [27-34]. Integrated photonic neurons based on Kerr nonlinearities have demonstrated all-optical activation functions [28], while active III-V semiconductor platforms have enabled reconfigurable nonlinear responses [29]. Although these approaches offer direct optical nonlinearity, they are often constrained by limited tunability, material complexity, or device-to-device variability [26, 35].

A second widely explored strategy employs optoelectronic nonlinearities, where optical signals are converted to the electronic domain to apply nonlinear functions and then re-encoded optically. Hybrid photonic-electronic neural networks exemplify this approach [36-38]. Nevertheless, repeated optical-electrical-optical conversions introduce latency and energy overhead, and scaling such systems to high-dimensional nonlinear operators remains challenging.

More recently, structural nonlinearity and optical phase encoding have emerged as an alternative paradigm [39-41]. In these approaches, nonlinear input-output relationships arise from purely linear optical propagation through complex structures, such as recurrent scattering systems [39], or diffractive networks [40-42], and related optical learning operators [43]. Free-space implementations have shown that linear optical elements, when combined with feedback or multiple scattering, can induce effective nonlinear processing. While structural nonlinearity is promising for various inference tasks, it suffers from a fundamental limitation: the resulting nonlinear functions are implicit and cannot be freely designed. Optical phase encoding is useful; however, its free-space configuration is bulky and limits its range of applications.

Taken together, existing photonic nonlinear computing approaches have not yet combined broad functional coverage, scalability, and hardware simplicity within a single integrated platform. In particular, no integrated photonics platform has yet demonstrated the capability to flexibly synthesize a broad class of nonlinear functions, ranging from high-order univariate functions to high-dimensional nonlinear operators.

In this work, we address this challenge by introducing a fundamentally different approach to nonlinear photonic computing. The key to our approach is to leverage random Fourier feature (RFF) mapping based on optical phase encoding. RFF mapping is a widely used technique for approximating shift-invariant positive definite kernels [44] and has been adopted for efficient attention mechanisms [45]. More importantly, it is recognized as an effective tool for representing high-frequency functions [46].

We here present an optical random encoder on a silicon photonic platform to implement RFF mapping and convert nonlinear function approximation in the input space into an equivalent linear computation in a randomized feature space, enabling efficient and scalable nonlinear function approximation. Guided by the RFF-based theoretical guarantee, we experimentally demonstrate that the proposed optical encoder can represent diverse nonlinear functions, including tenth-order Legendre polynomials, special functions (Voigt, Fermi–Dirac, and Fresnel), neural-network activation functions, two-dimensional (2D) nonlinear functions, and a 10-dimensional softmax operator. These results establish a practical strategy for nonlinear function approximation in integrated photonics, with potential applications extending beyond neural networks to scientific computing and signal processing.

## Results

### Optical encoder

The proposed optical encoder is designed to approximate a broad class of nonlinear functions $\boldsymbol{y} = \boldsymbol{f}(\boldsymbol{x}) \in \mathbb{R}^L$ for an input vector $\boldsymbol{x} \in \mathbb{R}^D$ using an optoelectronic implementation. For clarity, we first consider the scalar-output case $(L = 1)$ without loss of generality. The key idea is to represent the function $f(\boldsymbol{x}) \in \mathbb{R}^1$ in the Fourier domain and to approximate it using RFF as follows:

$$f(\boldsymbol{x}) = \int F_{\boldsymbol{\omega}} e^{i\boldsymbol{\omega}^{\mathrm{T}}x} d\omega \approx \sum_{\boldsymbol{\omega}\in\Omega} F_{\boldsymbol{\omega}} e^{i\boldsymbol{\omega}^{\mathrm{T}}\boldsymbol{x}}, \tag{1}$$

where $\boldsymbol{\omega} \in \mathbb{R}^D$ and $F_{\boldsymbol{\omega}} \in \mathbb{C}^1$ denote the frequency vector, randomly sampled from a $D$-dimensional set $\Omega$, and its corresponding Fourier coefficient, respectively. The superscript T denotes the transpose. RFF provides an efficient way to approximate arbitrary functions in an $L^2$ space by mapping the input into a randomized Fourier feature space, where nonlinear function evaluation can be reformulated as a linear combination of the basis functions $e^{i\boldsymbol{\omega}^{\mathrm{T}}x}$. This approach significantly reduces computational complexity and is powerful for representing high-frequency components [46].

To achieve the optical RFF mapping operation on a silicon photonic platform, we develop an optical encoder with $N$ input and $M$ output channels, integrated with phase modulators (PMs) and photodetectors (PDs), as shown in Fig. 1a. The operating principle is summarized as follows. An input signal $\boldsymbol{x} \in \mathbb{R}^D$ is encoded into an optical phase via a PM at each input channel; the electric field at the $j$-th input channel is represented as $\varphi_j(x) = \exp\,(i\boldsymbol{\alpha}_j^T\boldsymbol{x}) \in \mathbb{C}^1$ , where $\boldsymbol{\alpha}_j \in \mathbb{R}^D$ denotes the modulation strength at the $j$-th channel. The encoded signals are then mixed by a fixed random linear optical transformation due to multiple scattering in the optical encoder, the operation of which is represented by $M \times N$ complex-valued matrix $\boldsymbol{A}$. The signal at the $k$-th output channel is represented by $z_k = \sum_j^N A_{kj}\varphi_j(\boldsymbol{x}) \in \mathbb{C}^1$, $(k = 1 \ldots M)$. Square-law photodetection produces intensities as follows:

$$I_k = |z_k|^2 = \sum_{j=1}^{N}\sum_{j'=1}^{N} A_{kj}A_{kj'}^{*}\, e^{i\left(\alpha_j^{\mathrm{T}} - \alpha_{j'}^{\mathrm{T}}\right)x}. \tag{2}$$

The intensity $I_k$ serves as a nonlinear feature. A linear readout with the trainable weight vector $\boldsymbol{w} = (w_1, \cdots, w_M)^{\mathrm{T}} \in \mathbb{R}^M$ forms the output $y(x) \in \mathbb{R}^1$ as follows:

$$y(x) = \sum_{k=1}^{M} w_k I_k = \sum_{k=1}^{M}\sum_{j=1}^{N}\sum_{j'=1}^{N} w_k A_{kj}A_{kj'}^{*}\, e^{i\left(\alpha_j^{\mathrm{T}} - \alpha_{j'}^{\mathrm{T}}\right)x} = \sum_{j=1}^{N}\sum_{j'=1}^{N} F_{jj'} e^{i\left(\alpha_j^{\mathrm{T}} - \alpha_{j'}^{\mathrm{T}}\right)x}, \tag{3}$$

where $F_{jj'} = \sum_{k=1}^{M} w_k A_{kj} A_{kj'}^{*}$. Here, we define the set of effective frequencies as $\Omega = \{\boldsymbol{\omega} = \boldsymbol{\alpha}_j - \boldsymbol{\alpha}_{j'} | 1 \le j, j' \le N\}$. Using this notation, the output can be expressed as a finite Fourier series as follows:

$$y(x) = \sum_{\boldsymbol{\omega}\in\Omega} F_{\boldsymbol{\omega}} e^{i\boldsymbol{\omega}^{\mathrm{T}}\boldsymbol{x}}. \tag{4}$$

This representation corresponds to the RFF mapping, in which the initial frequencies $\boldsymbol{\alpha}_j$ are randomly drawn from a prescribed distribution and subsequently combined through the quadratic nonlinearity induced by photodetection. $F_{\boldsymbol{\omega}}$ is defined using the weight vector $\boldsymbol{w}$ to approximate the function $f(\boldsymbol{x})$ in a training phase.

Importantly, the effective frequencies $\boldsymbol{\omega} = \boldsymbol{\alpha}_j - \boldsymbol{\alpha}_{j'}$ arise from pairwise interactions of the encoded features. Therefore, the number of distinct frequency vectors $\boldsymbol{\omega}$ scales as $N^2$, provided that $\boldsymbol{\alpha}_j \neq \boldsymbol{\alpha}_{j'}$ for $j \neq j'$. This quadratic scaling, that is, $|\Omega| = O(N^2)$, in the spectral basis significantly enhances the expressivity of the model, enabling accurate approximation of high-frequency nonlinear functions. Meanwhile, since the output is obtained as a linear combination of the expanded features, the number of trainable weight parameters scales as $M$. According to theoretical analysis, the configuration guarantees uniform convergence with the squared approximation error scaling as $O(M^{-1})$.

We fabricated the optical encoder in a silicon photonics process. Figure 1b shows a micrograph of the full chip and Fig. 1c shows a close-up view of the optical encoder. The prototype device comprises $N = 32$ input PM channels, a passive optical encoder with a footprint of $33 \times 147\ \mu m^2$, and $M = 17$ PD outputs. Input laser light coupled into the chip is split into $N$ channels by a multi-mode interferometer tree and routed to the PMs. Multiple optical inputs are phase-modulated according to the input signals and then coupled into the interference region through the input waveguides. In the interference region, wave mixing is induced, depending on the relative phase differences among the inputs, as shown in Fig. 1d. The resulting interference pattern is read out as output intensities at the output waveguides. To promote richer interference among the inputs, the encoder was designed with arc-shaped input ports that concentrate interference near the centre of the mixing region and with dimensions that induce multiple reflections at the sidewalls.

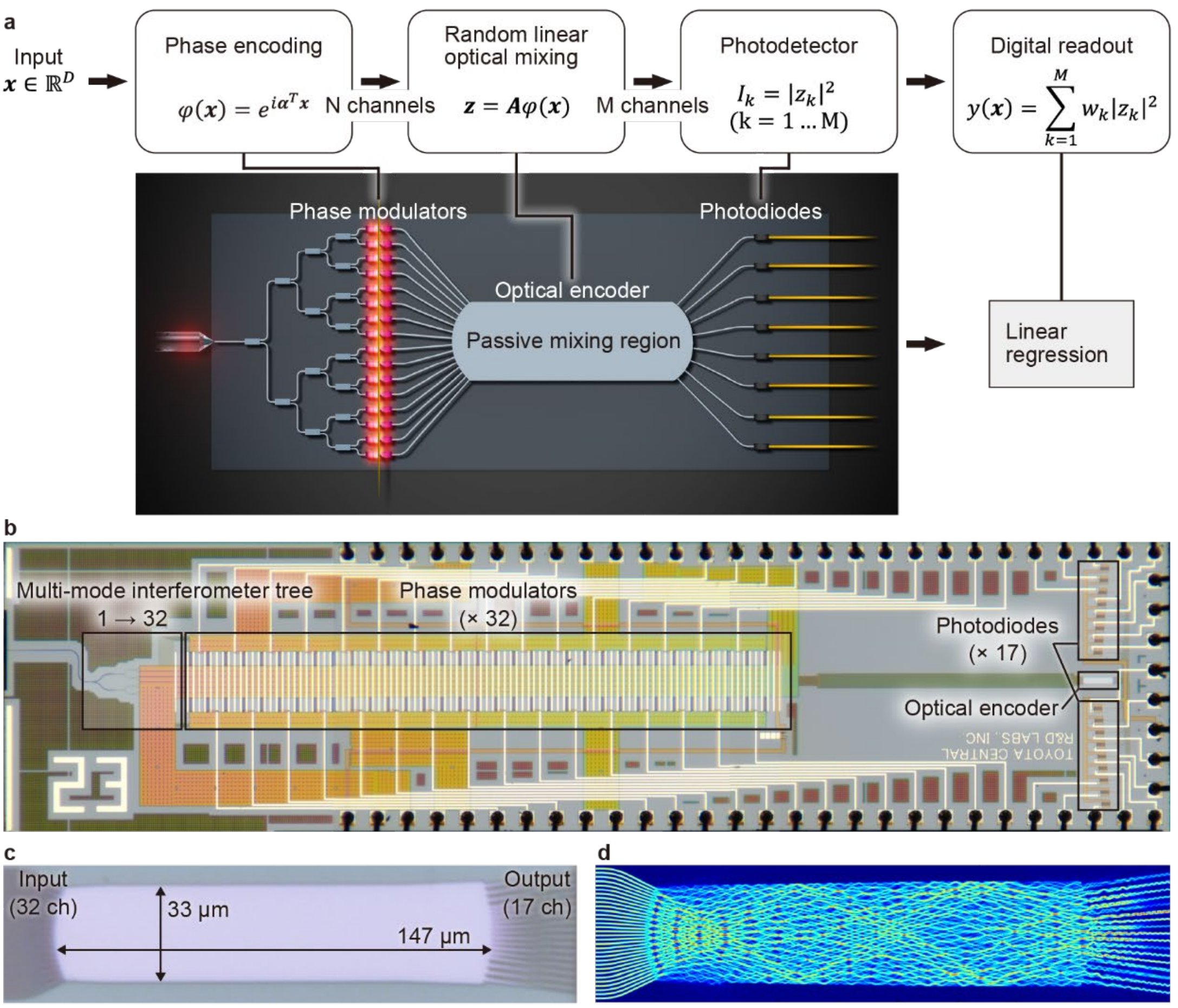

**Fig. 1: Optical RFF mapping operation (optical encoding) on silicon photonic platform.**

**a** Conceptual diagram of the proposed optical encoder implementing RFF. Input variables are encoded as optical phases, mixed by a passive photonic circuit, converted into nonlinear intensity features via square-law photodetection, and combined by a linear readout to produce the output. **b** Optical micrograph of the fabricated silicon photonic chip. **c** Close-up view of the optical encoder region showing the input waveguides, mixing region, and output waveguides. **d** Simulated optical field distribution inside the encoder (finite-difference time-domain), illustrating interference among phase-modulated inputs.

**Approximation of one-dimensional nonlinear benchmark functions**

We evaluate the function approximation capability of the optical encoder with a linear readout using one-dimensional nonlinear benchmark functions. The input phase parameter $\theta$ was swept from 0 to 2π in steps of 0.002π and a total of 1001 output data points were acquired. For each target function $\boldsymbol{f(x)}$, the regression labels were defined using $\boldsymbol{x}$ obtained by linearly mapping $\theta$ to the input domain of the target function (for example, $\boldsymbol{x} \in [-1, 1]$, depending on the function). The same $\theta$ value was applied to a subset of input channels of the encoder (tiled input), while the remaining channels were fixed at 0.000π.

Because the approximation accuracy depended on both the number and identity of the active input channels, we performed a preliminary screening procedure for each benchmark. Specifically, for each candidate active-channel count $ch_n \in \{1, 2, 4, 8, 16, 24, 32\}$, 100 distinct input-channel subsets were randomly sampled from the 32 available channels. For each sampled subset, the corresponding PD outputs were measured and evaluated using the regression protocol described in the Methods section. The active-channel count used for each benchmark was selected as the one yielding the lowest mean root-mean-squared error (RMSE) across the 100 sampled subsets.

Figures 2a and 2b show representative results for a sine wave and a sinc function. The predictions (blue markers) closely follow the reference curves (black lines), with RMSEs of 0.0555 and 0.0165, respectively. These results indicate that the encoder can approximate nonlinear functions accurately, with performance comparable to that of conventional bulky free-space optical systems [42, 43].

To evaluate performance across increasing levels of complexity, we next consider the regression of Legendre polynomials of orders 1–10. Figure 2c summarizes the dependence of RMSE and the coefficient of determination ($R^2$) on polynomial order. As the order increases, RMSE generally increases while $R^2$ decreases. Figures 2d and 2e show representative fits for orders 5 and 10, respectively. Although deviations appear near sharp extrema, the overall polynomial shapes are well reproduced. Even for the 10th-order polynomial, the target waveform remains clearly recognizable, demonstrating the applicability of the encoder to relatively high-order nonlinear functions.

As additional non-periodic benchmarks, we consider common neural-network activation functions. Figures 2f–2h show representative results for sigmoid, ReLU, and Swish functions, respectively. The corresponding normalized RMSE values are 0.0061, 0.0348, and 0.0327, respectively. These results suggest that the optical encoder can provide useful nonlinear features for activation mappings.

To further examine practical relevance, we consider three representative special functions, namely the Voigt profile [47-49], the Fermi–Dirac distribution [50], and the Fresnel integrals [51], which are frequently evaluated in spectroscopy, semiconductor physics, and wave optics. These functions are often computed repeatedly in simulations and numerical routines, making reductions in computational cost highly desirable. Figures 3a–3c summarize the regression results. For the Voigt profile, the encoder well agrees with the reference curves across multiple (σ, γ) settings (RMSE = 0.0026–0.0033), although larger deviations appear in the slowly varying tails ($|x| > 5.0$). For the Fermi–Dirac distribution, the predictions closely match the reference curves at moderate temperatures (RMSE = 0.0078–0.0146 for temperature $T_e$ = 300–1000 K), whereas larger deviations appear at lower temperatures (RMSE = 0.0382 at $T_e$ = 100 K and RMSE = 0.0698 at $T_e = 1\times10^{-9}$ K). For the Fresnel integrals, noticeable deviations are observed (C(x): RMSE = 0.0699; S(x): RMSE = 0.0650), particularly in highly oscillatory regions ($|x| > 0.25$).

Overall, the encoder achieves the highest accuracy for smooth and continuous targets, while accuracy decreases for functions with sharper transitions or stronger oscillatory behaviour. Nevertheless, the results demonstrate that the encoder can approximate a broad range of practically relevant one-dimensional nonlinear functions.

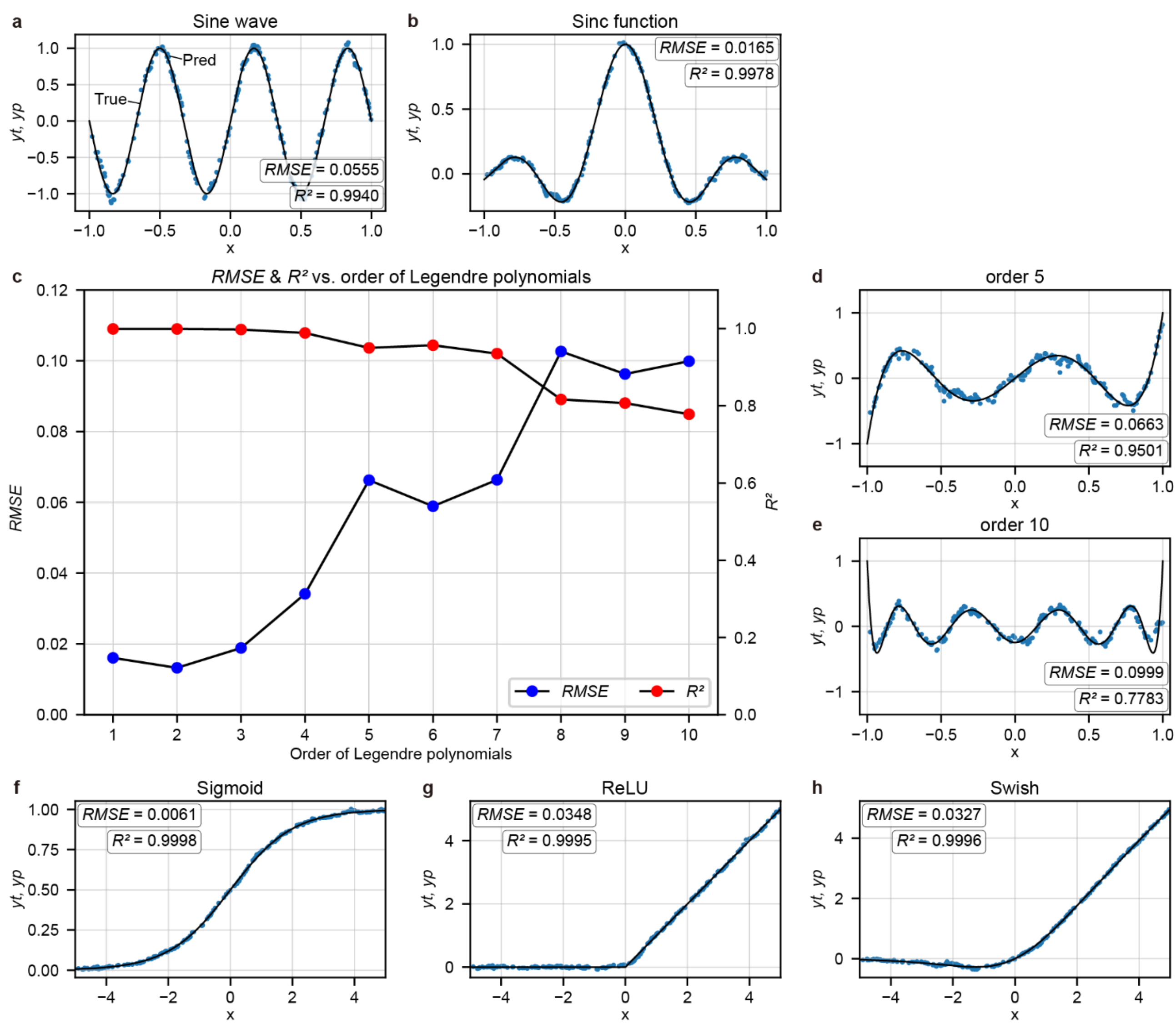

**Fig. 2: One-dimensional nonlinear function approximation.**

**a, b** Regression results for representative periodic functions (**a** sine and **b** sinc). Predictions closely follow the target functions across the input domain. **c** Dependence of RMSE and $R^2$ on the order of Legendre polynomials. **d, e** Representative fits for 5th- and 10th-order Legendre polynomials. **f–h** Regression results for neural-network activation functions (**f** sigmoid, **g** ReLU, and **h** Swish).

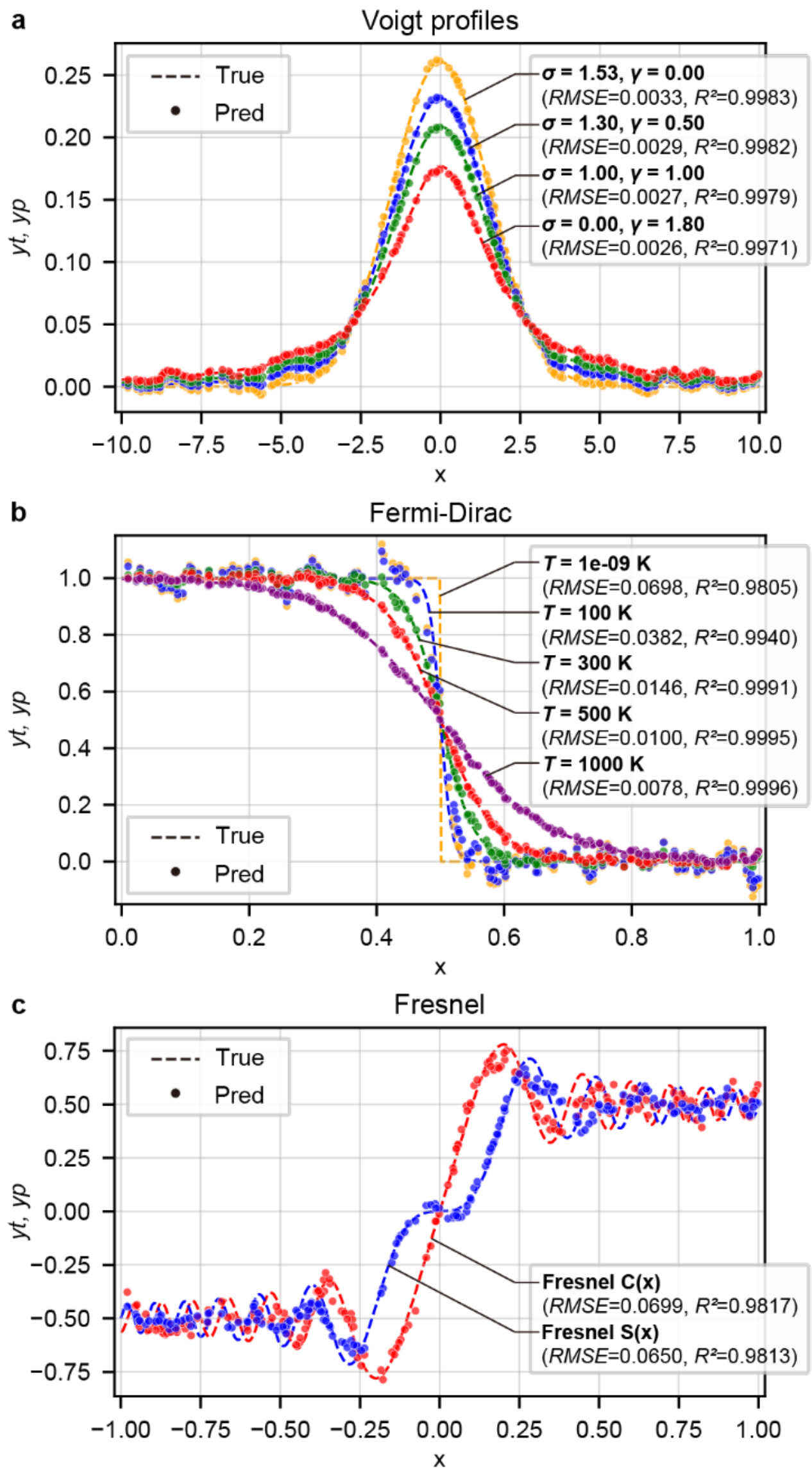

**Fig. 3: Approximation of scientific special functions.**

**a** Voigt profiles for various parameter (σ, γ) settings. **b** Fermi–Dirac distributions at various temperatures. **c** Fresnel integrals. The encoder reproduces the overall behaviour of the target functions. Deviations become more pronounced in highly oscillatory or flat regions.

**Approximation of two-dimensional nonlinear functions**

We next evaluate the optical encoder for approximating 2D nonlinear functions, $\boldsymbol{y} = \boldsymbol{f}(\boldsymbol{x_1}, \boldsymbol{x_2})$, to examine its ability to handle multivariate inputs. The input variables were sampled on a uniform grid over $\boldsymbol{x_1}, \boldsymbol{x_2} \in [-1, 1]$ and linearly mapped to phase drives $(\theta_1, \theta_2)$ in [0, 2π]. In the experiments, $\theta_1$ and $\theta_2$ were swept from 0.008π to 2π in steps of 0.008π, resulting in a uniform 250 × 250 grid (62,500 samples).

Because the regression accuracy depended on the input-channel configuration, we performed a preliminary screening procedure for each benchmark. Specifically, for each candidate tiling level $ch_n \in \{1, 2, \cdots, 15\}$, 100 distinct input-channel configurations were randomly sampled from the 32 available input channels. For each sampled configuration, the two phase inputs were assigned to $2 \times ch_n$ active channels in total, while all remaining channels were fixed at 0.000π. The corresponding PD outputs were evaluated using the five-fold cross-validation and out-of-fold prediction pipeline described in the Methods section. The tiling level used for each benchmark was selected as the one yielding the lowest mean RMSE across the 100 sampled configurations.

Figure 4 summarizes the regression results for four representative 2D functions, namely a 2D Gaussian, a quadratic form, a periodic function, and a radial sinc function. For visualization, the heatmaps shown in Fig. 4 were generated using the sampled configuration that yielded the lowest RMSE among the 100 configurations for the selected $ch_n$. The encoder achieves the highest accuracy for the smooth 2D Gaussian and quadratic form targets (RMSE = 0.0118 and 0.0494, respectively), whose heatmaps are dominated by low-spatial-frequency components. In contrast, the periodic and radial sinc functions produce larger errors (RMSE = 0.1597 and 0.0911, respectively) because they contain richer high-frequency content and sharper spatial variations, which are more difficult to represent with the present encoder and linear readout. Even in these more challenging cases, the predictions still capture the dominant global structures, such as the overall extrema pattern of the periodic function and the central lobe of the radial sinc function.

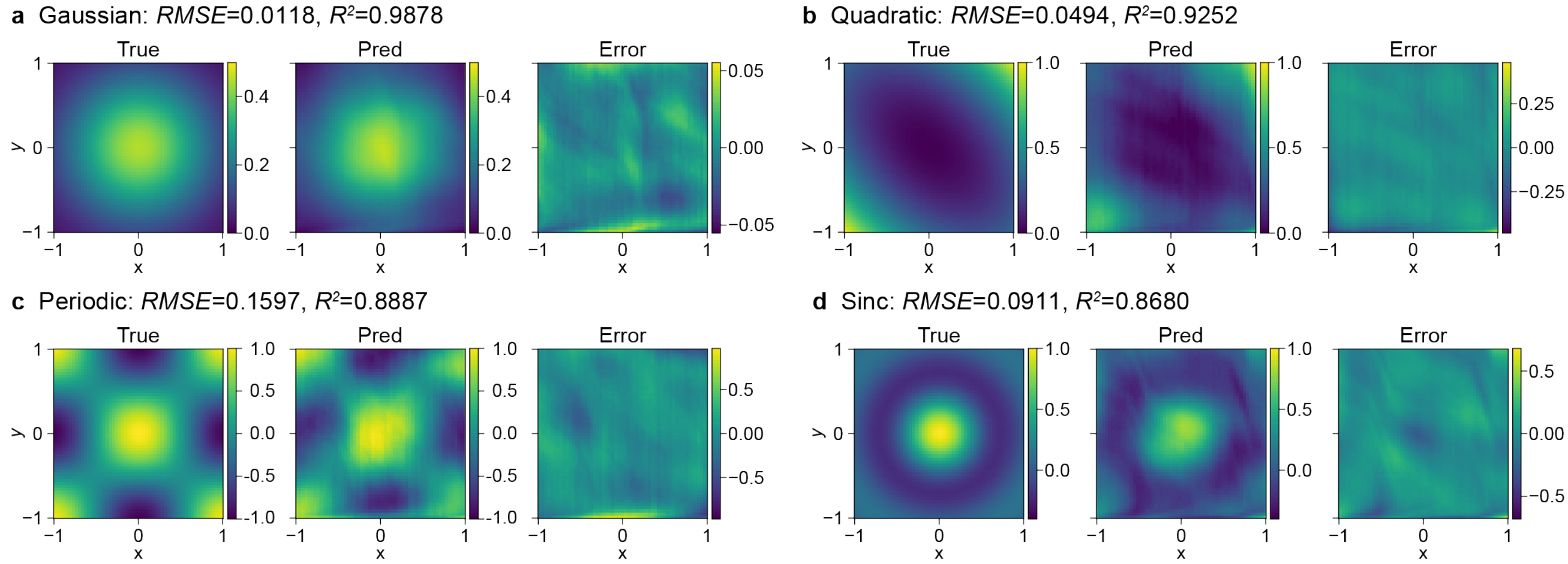

**Fig. 4: Two-dimensional nonlinear function approximation.**

Regression results for representative 2D functions. For each function, the ground truth, prediction, and residual (prediction minus ground truth) are shown. **a** 2D Gaussian, **b** quadratic form, **c** periodic function, and **d** radial sinc function. The encoder achieves higher accuracy for smooth functions. Functions with higher spatial-frequency components exhibit larger residuals.

**Approximation of softmax layer for practical application**

Softmax is widely used in classification networks, attention mechanisms, probabilistic inference, and large-scale foundation models, and is repeatedly applied across high-dimensional feature spaces. In transformer-based architectures in particular, it is a key component of the attention mechanism and directly affects both computational complexity and energy consumption. Despite its simple mathematical form, the softmax function involves exponential operations and global normalization, which incur significant computational and memory overhead. Hardware-efficient and parallelizable implementations are therefore of broad interest.

Motivated by the accurate approximation of activation functions (Figs. 2f–2h), we next investigate the optical encoder as a hardware-friendly approximation of a softmax layer in Fashion-MNIST classification. Figure 5a illustrates the evaluation pipeline. A multilayer perceptron (MLP) with two hidden layers (256 and 128 units, respectively) was trained using the standard cross-entropy loss to produce a 10-dimensional output vector (logits). After training, the logits were passed through a sigmoid function to generate intermediate vectors $\boldsymbol{z}$, which were used as inputs to both the digital softmax baseline and the proposed optical-encoder-based mapping. The sigmoid transformation was introduced to match the input range required by the optical encoder. For the baseline, the softmax function was directly applied to $\boldsymbol{z}$. For the proposed method, $\boldsymbol{z}$ was encoded into optical phase inputs, processed by the optical encoder, and mapped to a 10-dimensional output via a linear readout trained to approximate softmax($\boldsymbol{z}$). The linear readout was trained using the training dataset (60,000 samples) and performance was evaluated on the test dataset (10,000 samples). The approximation accuracy was quantified using the RMSE between the predicted output and the digital softmax output and classification accuracy was computed based on the argmax of the output vector.

Figure 5b shows box-and-whisker plots of the approximation error, quantified as the RMSE between the encoder output and the digital-softmax output, grouped by class. The maximum RMSE is below 0.010 and the median RMSE is below 0.005 for all classes, indicating high approximation accuracy. Figure 5c compares the resulting confusion matrices. The encoder with a linear readout achieves a classification accuracy of 0.8592, close to the digital-softmax baseline of 0.8747, and exhibits similar misclassification patterns across classes.

Although the approximation error remains small, a modest drop in classification accuracy ($\Delta = 0.0155$) is observed. Because the final prediction is determined by an argmax operation, even minor deviations in the output vector can alter the top-1 decision, particularly in near-tie cases. We attribute the accuracy gap primarily to such near-degenerate outputs, where sub-percent-level differences are sufficient to flip the predicted class.

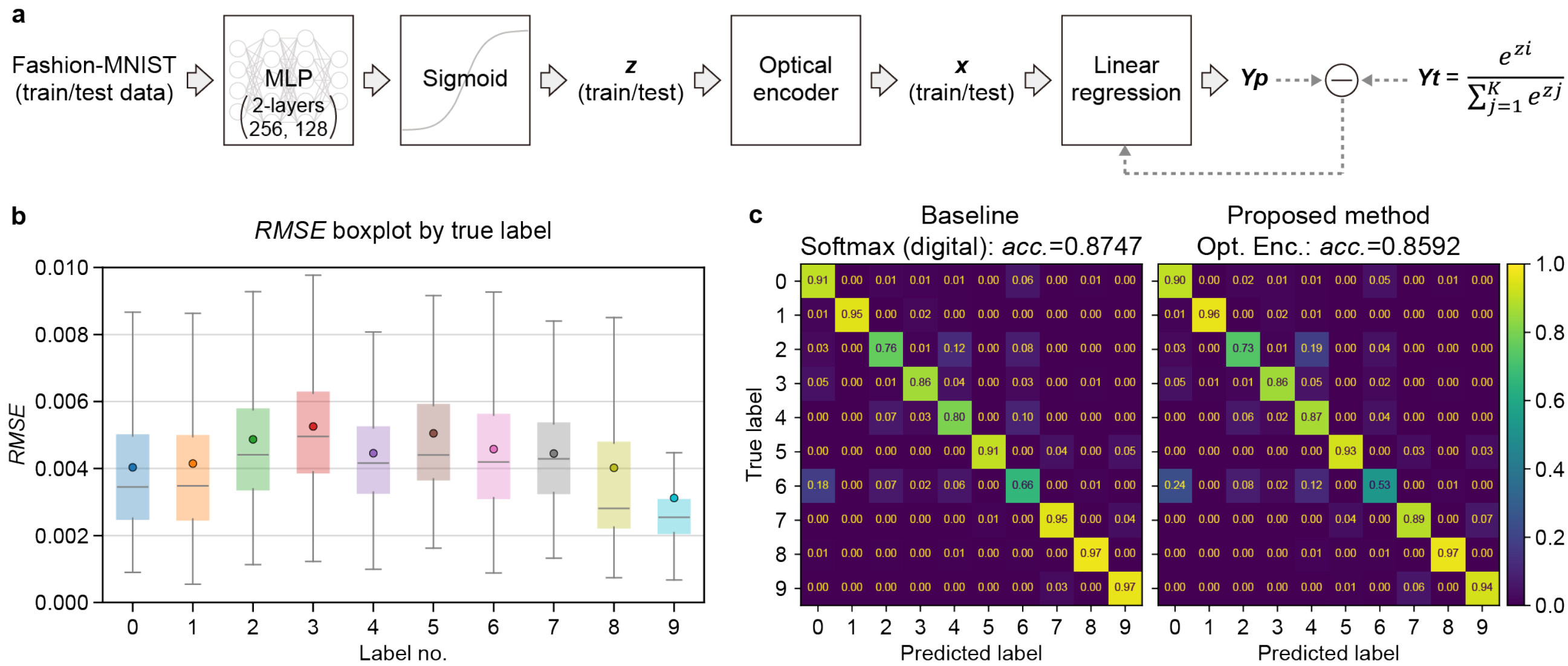

**Fig. 5: Softmax approximation for Fashion-MNIST classification.**

**a** Evaluation pipeline. A sigmoid layer was applied to the MLP output in order to match the encoder input range. **b** Boxplots of approximation error (RMSE) across Fashion-MNIST classes. **c** Confusion matrices obtained using the digital-softmax and the proposed approach. The overall classification patterns are preserved with a small reduction in accuracy.

**Parallelization: accuracy scaling with output count *M***

We next examine whether approximation accuracy can be improved by increasing the effective feature dimensionality. In a fully scaled hardware implementation, this would be accomplished by increasing the number of available PD outputs and/or operating multiple encoders in parallel. Realizing such a device-level scale-up, however, would require additional PMs, PDs, and fabrication effort. As a proof-of-principle, we therefore introduce a virtual parallelization scheme that emulates large-scale feature expansion using a single encoder.

Because the interference pattern is determined by the relative phase distribution across the 32 input channels, the same input signal can generate different output features when applied through different input-channel configurations. We therefore operate the same physical encoder sequentially under $P$ distinct input-channel configurations and concatenate the resulting outputs offline, thereby expanding the effective number of features from $M$ to $M_{eff} = P \times M$. The corresponding readout is written as

$$\hat{y}(x) = \sum_{p=1}^{P} \sum_{k=1}^{M} w_k{}^{(p)} I_k{}^{(p)}(x), \tag{5}$$

where $I_k^{(p)}(x)$ is the $k$-th PD intensity measured under the $p$-th input-channel pattern and $w_k^{(p)}$ is the associated trainable weight. (Experimental details are provided in the Methods section.) This procedure emulates the feature set of a larger system without requiring the fabrication of a larger photonic circuit. Because the same encoder is reused sequentially, however, the measurement cost increases approximately in proportion to $P$; accordingly, the present scheme is intended to isolate the benefit of feature-space expansion itself, rather than to demonstrate an immediate throughput advantage.

Figure 6a shows the dependence of the RMSE on the number of virtual encoder instances, $P$, for Legendre polynomial regression. For all polynomial orders, the RMSE decreases monotonically with increasing $P$. With $P$ = 20 virtual encoders (i.e., $M_{eff} = 20 \times 17$ output features), the RMSE falls below 0.02 for all orders. Figure 6b shows a representative result for the 10th-order polynomial. Compared with the single-encoder result in Fig. 2e, the prediction more faithfully reproduces the target waveform, with the RMSE improving from 0.0999 to 0.0150. These results indicate that increasing the available feature set is an effective route for improving approximation accuracy, consistent with the scaling behaviour discussed above.

We next apply the same virtual parallelization strategy to the scientific functions that exhibited the largest single-encoder errors in Fig. 3. Figure 7 shows RMSE trends as a function of $P$ together with representative fits obtained at $P = 20$. For the Voigt profile, the error improves by approximately one order of magnitude across all ($\sigma$, $\gamma$) settings, and the encoder well reproduces the reference curves even in the slowly varying tails. For the Fermi–Dirac distribution, the improvement is approximately one order of magnitude at moderate temperatures ($T_e$ = 300–1000 K), with the lowest-temperature case remaining challenging. For the Fresnel integrals, the RMSE decreases by approximately a factor of five, from 0.0699 and 0.0650 to 0.0110 and 0.0132 for $C(x)$ and $S(x)$, respectively, and the encoder well captures the reference curves even in the oscillatory regime ($|x| > 0.25$).

Overall, these results provide experimental evidence that expanding the effective feature dimensionality is a reliable route to higher approximation accuracy for this architecture. The present study should therefore be viewed as an experimental surrogate for future hardware scaling, rather than as the final large-scale implementation itself. Although this scheme increases the measurement time approximately in proportion to $P$, it establishes that large-scale feature expansion is a worthwhile design target. These findings motivate future device-level demonstrations based on physically parallel encoder architectures with increased numbers of outputs and/or encoder instances, in which the same scaling advantage could be realized without sequential measurement overhead.

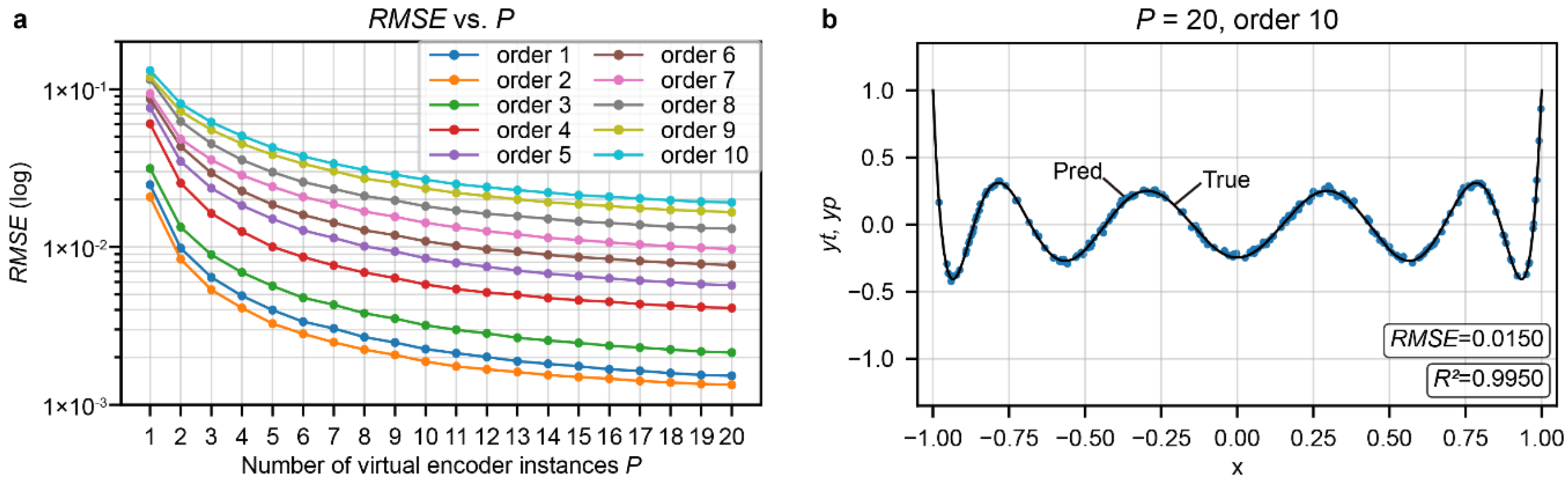

**Fig. 6: Accuracy scaling with virtual encoder instances for Legendre polynomial regression.**

**a** RMSE as a function of the number of virtual encoder instances for Legendre polynomial regression. Each curve corresponds to a different polynomial order. Markers indicate the mean RMSE over 500 random trials. **b** Representative result for the 10th-order polynomial using 20 virtual encoder instances, showing improved agreement with the target function compared to that achieved using a single encoder.

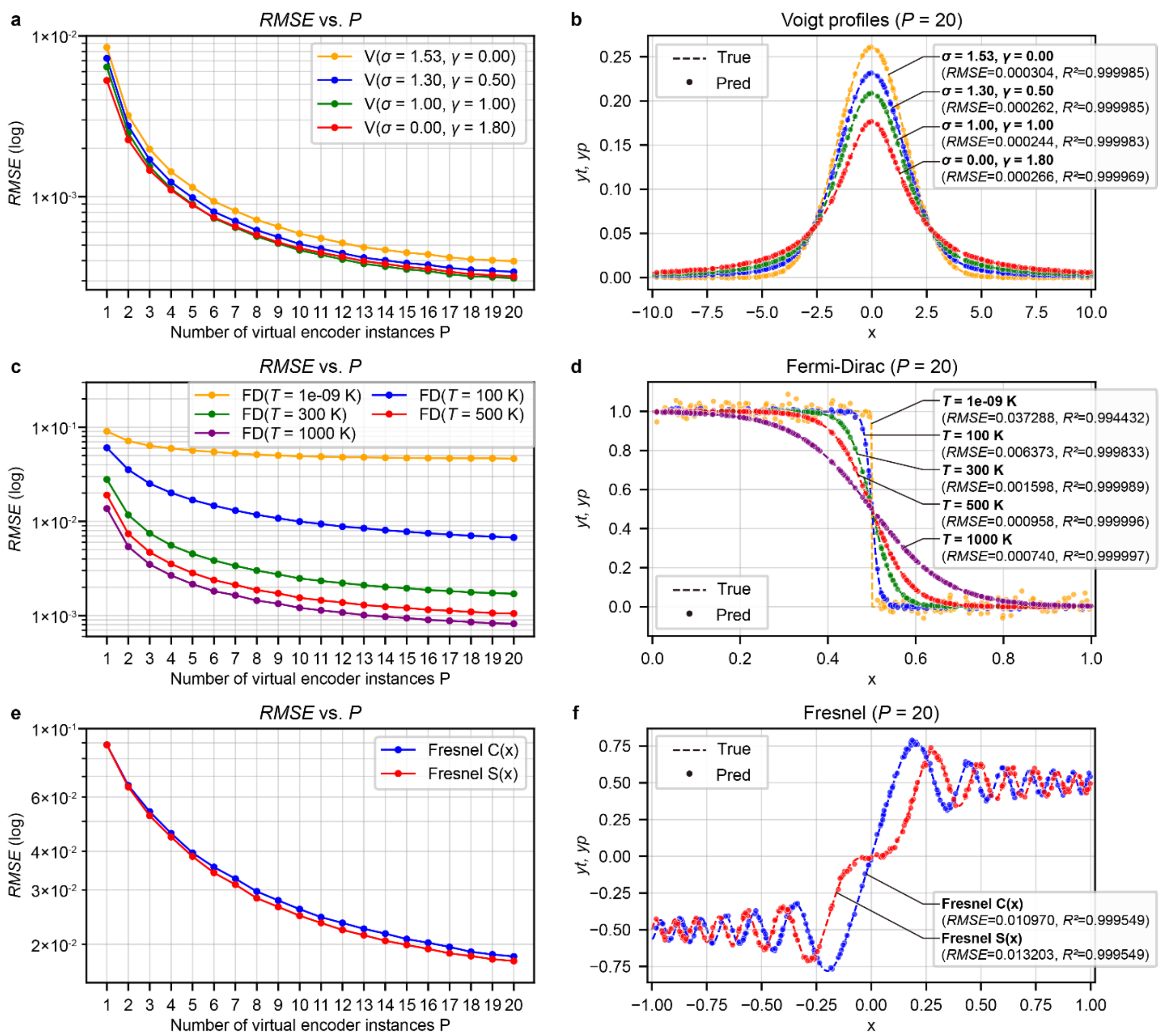

**Fig. 7: Accuracy scaling with virtual encoder instances for scientific special functions.**

RMSE as a function of the number of virtual encoder instances for representative special functions. **a,b** Voigt profiles, **c,d** Fermi–Dirac distributions, and **e,f** Fresnel integrals. Increasing the effective number of features improves approximation accuracy across all functions.

## Discussion

Here, we demonstrated an optoelectronic framework for nonlinear function approximation based on optical random feature encoding implemented in integrated photonics. Input signals are mapped into a high-dimensional feature space through optical phase encoding, linear interference, and square-law photodetection, and the target function is obtained via a trainable linear readout. This architecture enables nonlinear function approximation without requiring intrinsic optical nonlinear materials or repeated optical-electrical-optical cycles.

From a computational perspective, the proposed system can be interpreted as a physical implementation of random feature expansion, where nonlinear function approximation is achieved through linear regression in a randomized feature space. Unlike digital implementations of RFF, the feature mapping is performed in parallel by optical interference, eliminating the need for explicit computation of basis functions. This physical realization establishes a link between statistical learning theory and photonic hardware, and suggests a route toward hardware-accelerated nonlinear computation based on feature-space representations.

We demonstrated the approximation of a diverse set of nonlinear functions, including high-order polynomials, special functions, activation functions, and multidimensional mappings. The results show that the encoder achieves higher accuracy for smooth functions dominated by low-frequency components, while functions with strong oscillations or sharp transitions remain challenging. This behaviour can be attributed to the limited number of available features and the restricted set of effective frequency components represented by the encoder.

We further introduced a virtual parallelization strategy that increases the effective feature dimensionality without requiring additional physical hardware. By combining outputs obtained from multiple input-channel configurations, we emulated multiple encoder instances and expanded the available feature set. The observed monotonic reduction in approximation error with increasing effective feature number is consistent with random feature theory, which predicts error scaling on the order of $O({M_{eff}}^{-1})$. This result highlights feature dimensionality as a key resource governing approximation accuracy for this architecture.

Several limitations remain. First, the current system relies on offline training of the linear readout and real-time adaptation would require additional electronic or optoelectronic feedback mechanisms. Second, the accuracy is constrained by the number of available PD outputs and by the diversity of the effective feature set, which is determined by the optical mixing properties of the encoder. Third, although the proposed approach avoids complex optical nonlinear materials, it still relies on photodetection and electronic post-processing, and therefore does not constitute a fully optical nonlinear computing system.

In addition, while photonic hardware offers potential advantages in terms of bandwidth and parallelism, the present work does not provide a quantitative comparison in terms of energy efficiency or latency relative to digital implementations. In the current prototype, static power consumption in the PMs and peripheral electronics is non-negligible. System-level optimization will be necessary to fully realize the potential efficiency benefits. The present prototype employs thermo-optic PMs because of their simplicity, CMOS compatibility, and large phase tuning range with minimal optical loss [52,53]. However, their operation relies on resistive heating, resulting in non-negligible static power consumption (4.4 mW per 2π phase shift per element). This limitation is not fundamental to the proposed computing paradigm. Alternative phase modulation mechanisms, such as microelectromechanical systems [54,55], carrier-depletion-based modulators [56,57], or Pockels-effect-based platforms such as thin-film lithium niobate [58,59], have demonstrated substantially reduced power consumption, with especially low static holding power in microelectromechanical systems and Pockels-effect-based platforms. Replacing thermo-optic-based phase shifters with such technologies could substantially reduce the overall energy consumption, particularly for static phase holding, without altering the underlying computational framework.

Linear projections and repeated multiply–accumulate operations remain a major computational cost in many digital machine-learning pipelines. Because our encoder physically implements a fixed random mixing through optical interference, it may offload part of the feature generation process that would otherwise be computed electronically. Toward practical hybrid systems, co-design with electronic accelerators and robust photonic-electronic co-packaging, including chiplet-based integration, will be important next steps. A particularly promising direction is to integrate the optical encoder with a hardware-native readout engine, such as a silicon-photonics-based vector–matrix multiplication stage or other low-overhead analogue linear-projection hardware, rather than performing the readout entirely digitally [60]. Although the present work does not establish a quantitative benefit in terms of energy efficiency or latency, such a co-integrated architecture could in principle reduce the digital multiply–accumulate workload and lessen dependence on high-resolution, high-throughput A/D conversion.

Beyond system integration, several directions could further enhance the capability of this approach. Increasing the number of input channels and PD outputs would directly expand the accessible feature space. Designing optical encoders with tailored or controllable mixing properties could improve spectral coverage and approximation accuracy. On-chip training or adaptive calibration mechanisms could further improve robustness and flexibility. More broadly, extending this framework to higher-dimensional inputs and structured data may open opportunities in machine learning, scientific computing, and signal processing.

Overall, this work demonstrates that nonlinear function approximation can be achieved through linear photonic processing combined with physical random feature encoding. By shifting the burden of nonlinearity from device physics to feature-space representation, the proposed approach provides a practical route toward more capable photonic computing systems. The results suggest that photonic hardware can serve as a platform for implementing statistical learning primitives directly in the physical domain.

## Methods

### Experimental setup

Figure 8 shows the experimental optoelectronic setup. A continuous-wave laser at 1550 nm is coupled into the chip through a polarization controller and an optical attenuator. A multichannel DAC/ADC system is used to drive the phase modulators (PMs) and record the photodetector (PD) outputs. The present prototype employs thermo-optic PMs. To ensure stable operation, the chip temperature is actively controlled using a Peltier device maintained at 35 °C.

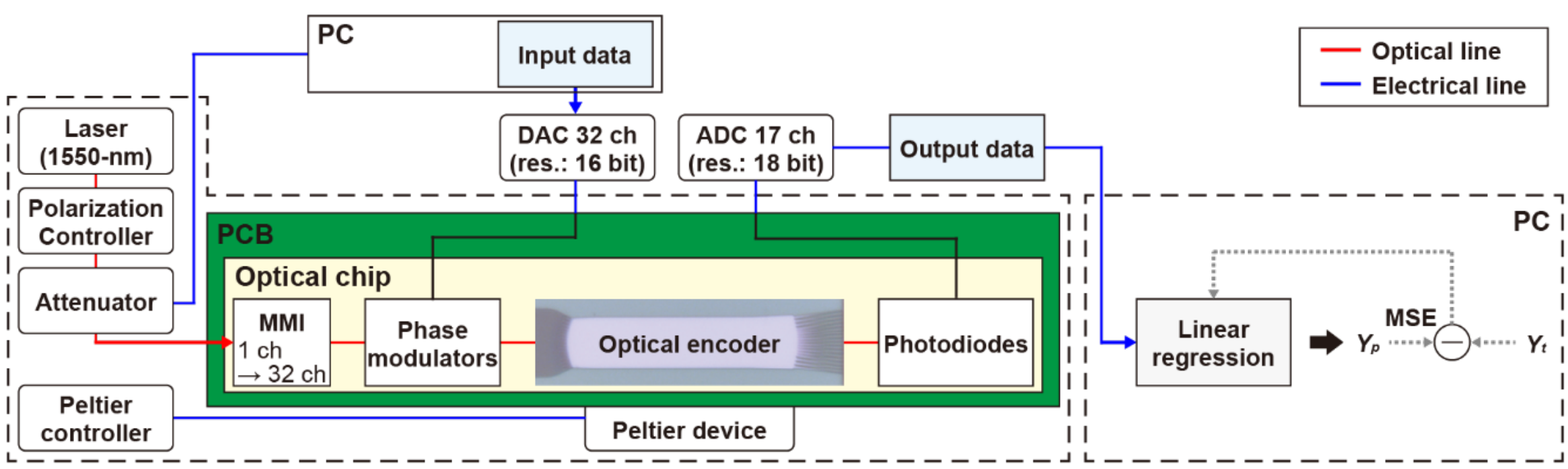

**Fig. 8: Experimental setup.**

Experimental optoelectronic setup for optical encoding and offline training, including a 1550-nm continuous-wave laser, polarization control and attenuation, a multichannel DAC/ADC for driving the PMs and recording the PD outputs, and digital linear regression. MMI: multi-mode interferometer.

### Offline linear regression and evaluation

In all regression experiments, the recorded PD intensity values were used as feature vectors, and the readout weights were trained offline by ordinary least-squares regression using scikit-learn (Linear Regression). Specifically, the linear model was fitted to minimize the mean squared error (MSE) between the predicted and target outputs. Regression performance was evaluated using the MSE, root-mean-squared error (RMSE), and coefficient of determination ($R^2$). Unless otherwise stated, the random seed was fixed at 42.

### Evaluation protocol for one-dimensional regression

For one-dimensional regression tasks, the input phase parameter $\theta$ was swept from 0 to $2\pi$ in steps of $0.002\pi$, yielding 1001 samples. For each benchmark, the scalar variable $x$ was obtained by linearly mapping $\theta$ onto the domain of the target function. The target values were then generated analytically from $x$ using the corresponding closed-form expression. The measured PD outputs were used as regression features.

To determine an appropriate tiling level, we screened the number of active input channels over $ch_n \in \{1, 2, 4, 8, 16, 24, 32\}$. For each candidate $ch_n$, 100 distinct subsets of active input channels were randomly sampled from the 32 available channels. For each sampled subset, the same scalar input $\theta$ was applied to all active channels, while the remaining channels were fixed at $0.000\pi$, and the corresponding PD outputs were recorded over the full $\theta$ sweep. This yielded 100 independently measured feature datasets for each $ch_n$.

Each dataset was evaluated independently using the same regression pipeline. Specifically, the dataset was randomly split into training and test subsets using an 80/20 split (test_size=0.2). A linear-regression model was trained on the training subset and evaluated on the held-out test subset, and the regression performance was quantified using MSE, RMSE, and $R^2$. The active-channel count used for each benchmark was selected as the one yielding the lowest mean RMSE across the 100 sampled subsets. For visualization, the representative prediction plots were generated using the sampled subset that yielded the lowest RMSE for the selected $ch_n$.

**Evaluation protocol for two-dimensional regression**

For two-dimensional regression tasks, the input coordinates $(\boldsymbol{x_1}, \boldsymbol{x_2})$ were sampled on a uniform 250 × 250 grid over [-1, 1] × [-1, 1], yielding 62,500 samples in total. These coordinates were linearly mapped to phase drives and applied to the selected input-channel configuration, while the remaining channels were held at fixed bias values. The recorded PD outputs were used as regression features. All two-dimensional target functions, including the Gaussian, quadratic, periodic, and radial sinc functions, were evaluated using the same experimental pipeline, with only the analytical form of the target function changed.

To determine an appropriate input tiling level, we screened $ch_n = 1, 2, \cdots, 15$. For each candidate $ch_n$, 100 distinct input-channel configurations were randomly sampled from the 32 available input channels. For each sampled configuration, the two phase inputs were assigned to $2 \times ch_n$ active channels in total ($ch_n$ channels per input variable), while all remaining channels were fixed at 0.000π. The corresponding PD outputs were recorded over the full input grid, yielding 100 independently measured feature datasets for each $ch_n$.

Each sampled configuration was evaluated independently using five-fold cross-validation (KFold, n_splits = 5, shuffle = True). In each fold, the linear-regression model was trained on the training subset and used to predict the held-out subset. Out-of-fold predictions were then assembled over all samples, ensuring that each data point was evaluated by a model that was not trained on that point. The final MSE, RMSE, and $R^2$ values were computed from the full set of out-of-fold predictions.

For each benchmark, the candidate $ch_n$ values were compared using the mean RMSE across the 100 sampled configurations and the tiling level used in the reported results was selected accordingly. For visualization, the prediction heatmaps were generated using the sampled configuration that yielded the lowest RMSE among the 100 configurations for the selected $ch_n$.

**Evaluation protocol for softmax mapping**

To evaluate the approximation of a softmax mapping, we used the Fashion-MNIST dataset, which consists of 60,000 training samples and 10,000 test samples across 10 classes. A multilayer perceptron (MLP) was trained to generate input vectors for the softmax approximation task.

The MLP consisted of two hidden layers with 256 and 128 units, respectively, followed by a 10-dimensional output layer. The model was trained using the cross-entropy loss with the Adam optimizer (learning rate = $1 \times 10^{-3}$) for 10 epochs and a batch size of 128. The choice of this MLP architecture is not critical; a lightweight network was adopted to generate representative feature vectors while keeping the computational cost low. After training, the output logits of the MLP were transformed using a sigmoid function to obtain bounded input vectors $z \in [0, 1]^{10}$. These vectors were used as inputs to both the digital softmax baseline and the optical-encoder-based mapping.

For the baseline, the softmax function was applied directly to $z$. For the proposed method, the vector $z$ was encoded into optical phase inputs, processed by the optical encoder, and mapped to a 10-dimensional output via a linear readout. The linear readout was trained to minimize the MSE between the predicted output and the reference softmax($z$).

The linear regression model was trained using the full training dataset (60,000 samples) and evaluation was performed on the test dataset (10,000 samples). Classification accuracy was computed using the argmax of the output vector. In addition, approximation accuracy was evaluated using the RMSE between the predicted outputs and the corresponding softmax outputs.

**Experimental method for proposed virtual parallelization**

To emulate an increase in the number of encoder instances without additional hardware, we implement a virtual parallelization scheme based on feature concatenation across multiple input-channel configurations. In this approach, the same physical encoder is operated repeatedly with different input-channel patterns, and the resulting outputs are concatenated to form an expanded feature vector.

Figure 9 shows the experimental method of the proposed virtual parallelization. Specifically, for each virtual encoder instance, a distinct subset of input channels is selected from the available inputs and the corresponding PD intensity outputs are recorded. The outputs obtained from $P$ such configurations are concatenated offline, resulting in an effective feature dimensionality of $M_{eff} = P \times M$, where $M$ is the number of physical detector outputs. This procedure enables a virtual expansion of the feature space without increasing the number of physical PDs.

To evaluate the dependence of approximation accuracy on the effective number of encoder instances, we generated multiple combinations of input-channel configurations using random sampling. For each target number of virtual encoders, multiple independent trials (500 trials per $P$ value) were performed by randomly sampling distinct subsets of input-channel configurations. In each trial, the concatenated feature matrix was used to train a linear regression model under a fixed train/test split (test size = 0.2), and the regression performance was evaluated using RMSE. For each $P$ value, the reported RMSE values were obtained by aggregating results over 500 trials; the mean RMSE values were computed across trials. In addition, for visualization, the representative prediction results were obtained from the trial that achieved the lowest average RMSE across all target functions.

This virtual parallelization approach enables systematic evaluation of how the effective feature dimensionality affects approximation accuracy. The observed monotonic decrease in RMSE with increasing number of virtual encoder instances is consistent with the expected scaling behaviour of random feature models, where the approximation error decreases as the number of features increases.

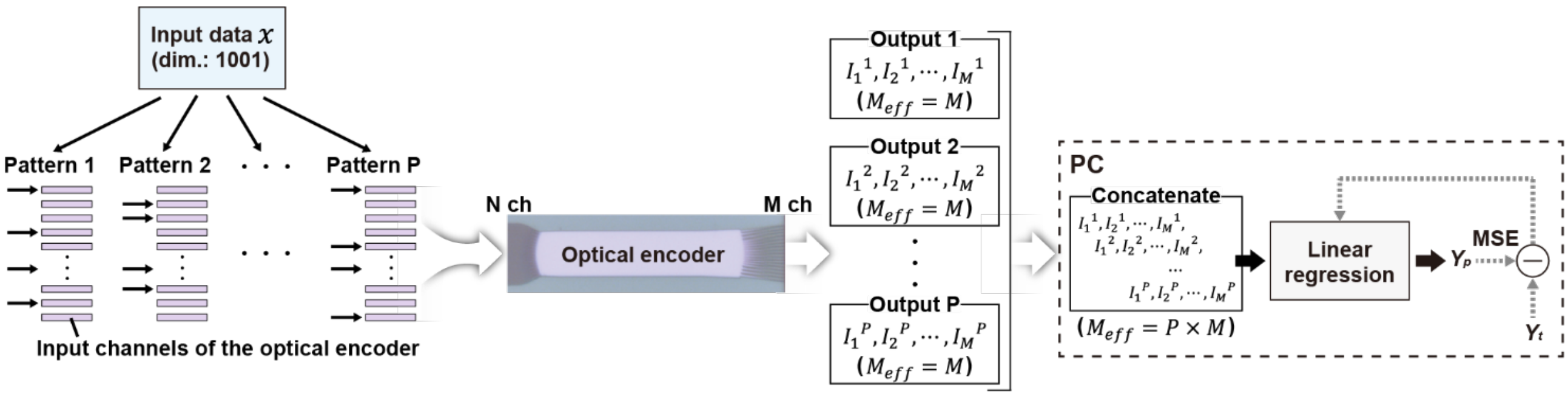

**Fig. 9: Experimental method for virtual parallelization.**

Schematic illustration of the virtual parallelization approach. A single optical encoder is operated multiple times with different input-channel configurations, producing distinct feature vectors for the same input. These features are concatenated to form an expanded representation, which is used for linear regression. Multiple configuration sets are randomly sampled to evaluate performance across trials. This procedure emulates multiple encoder instances without increasing hardware complexity and enables systematic investigation of accuracy scaling with feature dimensionality.

**Data availability**

The data that support the findings in this paper are provided in the main text. Additional data related to this study can be made available from the corresponding authors upon request.

**Acknowledgments**

The authors thank FORTE Science Communications (https://www.forte-science.co.jp/) for English language editing. The authors also acknowledge the use of ChatGPT (OpenAI) and Copilot (Microsoft) during the preparation of this manuscript for assistance with English grammar, phrasing, and Japanese-to-English translation. All AI-assisted content was reviewed and edited by the authors, who take full responsibility for the final content of the manuscript.

**Author contributions**

Ay.M. constructed the experimental system, conducted the experiments, performed the evaluations, prepared all the figures, and wrote the manuscript. I.T. and M.N. conceived the concept of the silicon photonics-based optical encoder. M.N. designed the optical encoder and set up the measurement system for the silicon photonics chip. S.S. proposed the application of the optical encoder to nonlinear function prediction, performed data analysis, theoretically clarified the operating principle of the optical encoder, and contributed to writing the manuscript. Ay.M., I.T., At.M., R.K., and S.S. discussed the experimental results, contributed to the interpretation of the data, and supervised the research. All authors read and approved the final manuscript.

**Competing interests**

The authors declare no competing interests.